\documentclass[aps,prl,twocolumn,groupedaddress,dvips]{revtex4}
\usepackage[dvips]{graphicx}
\begin{document}
\title{Spin-Polarized Electron Transport through Nanometer-Scale Al Grains}
\author{L. Y. Zhang, C. Y. Wang, Y. G. Wei, X. Y. Liu, and D. Davidovi\'c}
\affiliation{Georgia Institute of Technology, Atlanta, GA 30332}
\date{\today}
\begin{abstract}
We investigate spin-polarized electron tunnelling through
ensembles of nanometer scale Al grains embedded between two
Co-reservoirs at 4.2K, and observe tunnelling-magnetoresistance
(TMR) and effects from spin-precession in the perpendicular
applied magnetic field (the Hanle effect). The spin-coherence time
($T_2^\star$) measured using the Hanle effect is of order $ns$.
The dephasing is attributed to electron spin-precession in local
magnetic fields. Dephasing process does not destroy $TMR$, which
is strongly asymmetric with bias voltage. The asymmetric TMR is
explained by spin relaxation in Al grains and asymmetric electron
dwell times.
\end{abstract}

\pacs{73.23.-b,73.63.-b,73.21.-b}
\maketitle

Long spin relaxation times for polarized carriers are necessary
for development of spintronic
devices.
In quantum dots, spin relaxation times are strongly enhanced
compared to bulk, and the spin of an electron confined in a
quantum dot is a candidate quantum bit~\cite{loss,burkard}.
Unfortunately, the spin-coherence time $T_2^\star$ measured in a
semiconducting quantum dot is only $\sim ns$,~\cite{johnson1}
despite the fact that the spin-relaxation time ($T_1$) is
extremely long, up to a $\sim ms$~\cite{fujisawa,elzerman}. In
GaAs quantum dots, dephasing is caused by spin precession around
an effective magnetic field created by nuclear spins.

Spin-injection and detection from ferromagnetic electron
reservoirs is a well-known technique to measure spin relaxation
time~\cite{johnson}. In this paper we study electron
spin-injection in tunnelling junctions containing a large number
of embedded Al-grains.

\begin{figure}
\includegraphics[width=0.47\textwidth]{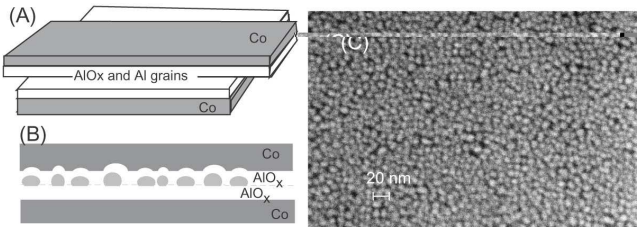}
\caption{A: Sketch of the tunnelling device geometry. B: Sketch of
the tunnelling junction cross-section. C: Scanning electron
micrograph of Al grains.~\label{fig1}}
\end{figure}

Our tunnelling device is sketched in Fig.~\ref{fig1}-A. The top
and the bottom layers are 100\AA\ thick Co films. The width/length
is 1.5mm/15mm and 1mm/20mm for the bottom and the top layer,
respectively. The sample cross-section, sketched in
Fig.~\ref{fig1}-B, shows nanometer scale Al grains embedded in
aluminum oxide. The device is a recreation of a tunnelling device
made by Zeller and Giaver in the 1960s~\cite{giaver}, which
demonstrated Coulomb Blockade for the first time. The difference
between our sample and the prior devices is that we have
spin-polarized leads.

The device is fabricated in two evaporation steps. First, we
thermally evaporate a Co film on a $SiO_2$ substrate, through a
mask at $4\cdot 10^{-7}$ Torr pressure. The deposition of cobalt
is stopped by closing the shutter. Next, we change the metal
source to Al and evaporate Aluminum in high vacuum, while the
shutter remains closed. Then we open the shutter for 1 second and
close the shutter again. The deposition rate is 0.2nm/sec. So, Co
layer is now covered with a seed layer of Al with nominal
thickness 0.2nm.

Our next deposition step is reactive deposition of aluminum oxide.
In this step, oxygen is introduced into the chamber, which exposes
the cobalt surface to oxygen vapor. The oxidation of Cobalt
surface should be minimal, because paramagnetic impurities in
cobalt-oxide could affect spin-polarized tunnelling. Our  strategy
to minimize oxidation of  Co is to apply as little oxygen as
possible for as little time as possible. In addition, the seed
layer also provides some protection of Co before passivation by
the deposited aluminum oxide. The seed layer must be very thin,
because if any metallic Al remains on Co surface after oxidation,
then spin-polarized tunnelling density of states will be reduced.

Immediately after closing the shutter the second time, we
introduce oxygen into the deposition chamber while evaporating Al.
The oxygen is introduced at a flow rate of 200\, sccm. The chamber
is continuously pumped with the cryopump gate valve fully open.
Oxygen pressure increases and reaches $10^{-5}$ scale Torr in few
seconds and stabilizes at $3\cdot 10^{-5}$ Torr after 30 sec. Only
during these initial 30 seconds, while the pressure increases and
stabilizes, cobalt surface with a 0.2nm seed layer of Al is
exposed to oxygen. After 30 sec, when the pressure is stabilized,
we open the shutter and evaporate 5nm of Al at a rate of 0.2nm/s,
to deposit the bottom aluminum-oxide film, which is 7nm thick.

In general, the thickness of the deposited oxide at fixed aluminum
evaporation rate will be an increasing function of oxygen
pressure. In our case, thickness of the deposited oxide versus
pressure saturates at 7nm at approximately $1\cdot 10^{-5}$ Torr.
Any further increase in oxygen pressure will not increase the
aluminum oxide thickness. Consequently, in our deposition process
nearly all Al atoms that are deposited at $3\cdot 10^{-5}$ Torr
are oxidized, however, the oxygen pressure is only three times
larger than the minimal oxygen pressure for the oxidation of Al
(the saturation pressure).

The oxygen pressure of $3\cdot 10^{-5}$ is substantially smaller
than typical oxygen pressures used to thermally oxidize Al
surfaces in tunnelling junctions. For example, in
Ref.~\cite{ralph}, nanometer sized Al nanoparticles were oxidized
at 0.1 Torr of oxygen for 1-2 minutes. This process created
tunnelling barriers of resistance in $M\Omega$ range, which
corresponds to oxide thickness of approximately 1nm. Since our
oxygen pressure is smaller by four orders of magnitude and the
oxidation time is shorter, the thickness of the surface aluminum
oxide in our case must be considerably smaller than 1nm. Thus, we
expect that the seed Al-layer of nominal thickness 0.2nm provides
some protection of Co surface from oxidation.

Prior to this work, this reactive evaporation technique was used
to create tunnelling junctions containing a single metallic
grain.~\cite{drago1} The junctions were of high quality and they
displayed well resolved Coulomb-Blockade steps and discrete energy
levels of the grain at low temperatures. So, the aluminum oxide in
our samples is a suitable insulator for the studies of properties
of metallic grains.

The sample, which is now passivated, is exposed to air and the
mask is replaced. Next, the sample is evacuated to base pressure
and we deposit 1.5 nm of Al, which creates isolated grains, as
shown by the image in Fig.~\ref{fig1}-C. Then we deposit another
layer of aluminum oxide, by evaporating 5nm of Al at rate 0.2nm/s
at $3\cdot 10^{-5}$ Torr of oxygen. Finally, we deposit the top Co
layer.

The average grain diameter is $\sim 6 nm$. If we assume that the
grains are hemispherical, analogous to Ref. ~\cite{ralph}, we
estimate that the average electron-in-a-box mean level spacing is
$0.2meV$.  Note that there is a relatively wide distribution of
grain diameters in Fig.~\ref{fig1}-C, as some grains have
coalesced. Hence, the range of level spacings in the ensemble is
rather large.

In addition, the grains are exposed to oxygen vapor before
deposition of the top oxide layer, at $3\cdot 10^{-5}$ Torr for 30
seconds. As a result, the grain surface is oxidized from above,
but we expect that the oxide thickness is considerably smaller
than 1nm, as discussed above. Additionally, there is generally
chemisorbed oxygen remaining on the underlying oxide surface.
Consequently, the grain surface may be oxidized from below. Thus,
the average size of the metallic core of the grains could be
smaller than the apparent grain size because of this effect, by up
to about 1 nm.

The number of grains in the junctions is $N=2.5\cdot 10^{10}$.
Although the junctions are very large, the resistance of the
junctions ($R$) varies significantly among samples made at the
same time. $R$ is in the range $1k\Omega<R<1M\Omega$. We also make
tunnelling junctions as described above but without Al grains and
find these devices to be insulating. In addition, we make control
samples without Al grains and with the aluminum-oxide layers at
half the thicknesses from those above. The control sample
resistance is in the same range ($1k\Omega<R<1M\Omega$), which
shows that tunnelling in the devices with grains take place via
the grains.

The fluctuations in sample resistance among samples made at the
same time show that the tunnel current must be dominated by the
current flow through weak spots. Consequently, the number of
grains that are active in transport is $\ll N$. The weak spots may
result from thickness variations in the oxide layer across the
junction area or from defects in the oxide, or from both.

We measured the surface roughness of a single aluminum oxide film
deposited over $SiO_2$ by the atomic-force microscope and found
that it was $\approx nm$. This surface roughness can cause weak
spots in the tunnelling barrier, because the tunnel resistance
decay length in oxides (~0.1nm) is much smaller than the surface
roughness.

In addition, it is known that amorphous aluminum oxide has
coordination number defects, which may be caused by oxygen
vacancies.~\cite{zhao} These defects could give rise to hole traps
near the valence band edge, which could result in weak spots for
tunnelling. These oxygen vacancies could be paramagnetic, which
could affect spin-polarized tunnelling.

\begin{figure}
\begin{center}
\includegraphics[width=0.47\textwidth]{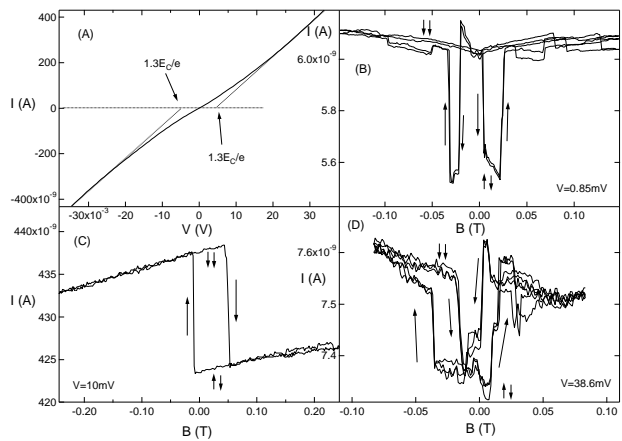}
\caption{A: IV-curves of sample 1  at 4.2K. B, C, and D: Current
versus parallel applied magnetic field at 4.2K, in samples 1,2,
and 3, respectively.~\label{fig2}}
\end{center}
\end{figure}

In this paper we present three devices with Al grains.
Fig.~\ref{fig2}-A displays the IV-curve of sample 1 at 4.2K. The
other two samples have similar IV-curves. The conductance is
suppressed at zero bias voltage, as expected from Coulomb-Blockade
on Al-grains, consistent with Ref.~\cite{giaver}.

The average charging energy $E_C$ is obtained by extrapolating the
linear part of the IV-curve at high V to zero I and finding the
offset voltage, as indicated. We averaged the Coulomb Staircase
IV-curve~\cite{averin2} over the background charge, capacitance
ratios, and over capacitance range $(C/4,7C/4)$, where $C$ is the
total capacitance of a grain. The corresponding offset was
$1.3E_C/e$, where $E_C=e^2/2C$.  $E_C$ is $4.3meV$ in samples 1
and 2 and $6meV$ in sample 3.


Figs.~\ref{fig2}-B,C, and D display current versus magnetic field
in samples 1,2, and 3 respectively, at constant bias voltage. All
samples exhibit $TMR$, which demonstrates that the tunnelling
current is spin-polarized. Samples 1 and 3 exhibit a spin-valve
effect: at a large negative field, the magnetizations of the
Cobalt reservoirs are down. If the field increases, the
magnetization of one Cobalt reservoir switches direction, and the
tunnelling current drops abruptly from $I_{\downarrow\downarrow}$
to $I_{\downarrow\uparrow}$. Finally, at a larger positive field
the current jumps back up to $I_{\uparrow\uparrow}\approx
I_{\downarrow\downarrow}$.

In addition to these abrupt transitions in TMR, we find that $TMR$
varies continuously with the magnetic field and it fully saturates
in the applied field of $\sim 1T$. Co-films are generally
multi-domain, and the average domain size in Co films is of the
order of 1 micron.~\cite{unguris}  If many domains were involved
in providing the TMR signal, one would expect the resistance
transitions to be gradual due to the spread in coercive fields
from domain to domain. Thus, only a portion of the sample of order
domain size or less is responsible for the abrupt TMR transitions.
This behavior is in agreement with the finding that the tunnelling
current is dominated by weak spots.

In sample 2, there is only one jump near zero field, followed by a
broad $TMR$ background that saturates  at $B\sim 1T$, which shows
that only one cobalt electrode exhibits an abrupt transition with
magnetic field. The abruptness of the transition indicates again
that this sample is sensitive to a very small fraction of the
physical device. However, in contrasts to samples 1 and 3, the
magnetic behavior of Co on one side of the effective contact area
indicates the presence of a very persistent magnetic defect, which
could be for example a 360 degree domain wall.~\cite{hubert}

Although our junctions are not ideal, we can learn about the
physics of spin-polarized tunnelling through grains by studying
how the abrupt resistance transitions depend on bias voltage and
the perpendicular applied magnetic field. TMR corresponding to
these transitions is a measure of the spin-polarization of the
current. The number of particles that contribute to the abrupt
transitions is very small. It is certainly smaller than the number
of particles that fit under a micron scale domain in Cobalt (
roughly $10^4$). The abrupt TMR transitions are reproducible when
the magnetic field cycle is repeated, as seen in Fig.~\ref{fig2}.
For the Hanle effect studies, we select devices that exhibit
spin-valve effect.

\begin{figure}
\includegraphics[width=0.47\textwidth]{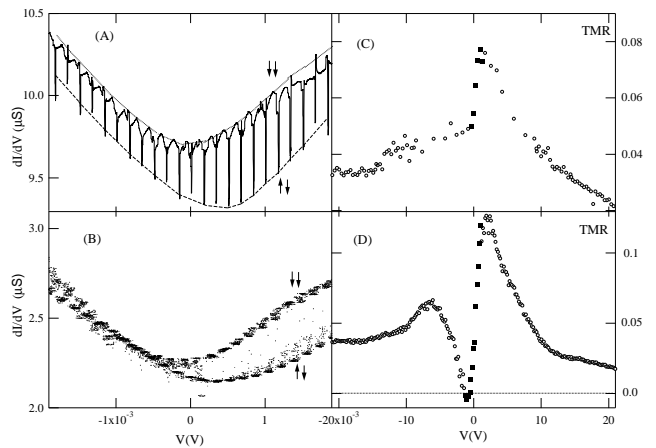}
\caption{A and B: Differential conductance versus bias voltage in
samples 1 and 2, respectively. C and D: $TMR$ versus bias voltage
in samples 1 and 2, respectively.~\label{fig3}}
\end{figure}

The tunnelling magnetoresistance is calculated as
\begin{equation}
TMR=(I_{\uparrow\uparrow}-I_{\uparrow\downarrow})/I_{\uparrow\downarrow},
\label{tmr}
\end{equation} where the current values were taken immediately before
and after the resistance transitions. Figs.~\ref{fig3}-A and B
display differential conductance $G$ with bias voltage in samples
1 and 2, respectively. $G$ is measured by the lock-in technique.
As the bias voltage is varied slowly at 3mv/hr, the magnetic field
is swept between -0.25T and 0.25T at 0.01Hz. The differential
conductance switches between $G_{\uparrow\uparrow}$ and
$G_{\uparrow\downarrow}$ when the magnetizations switch alignment.

$G_{\uparrow\uparrow}-G_{\uparrow\downarrow}$ changes
significantly when the bias voltage varies in a narrow interval
around zero-bias voltage.
In sample 2, the asymmetry in
$G_{\uparrow\uparrow}-G_{\uparrow\downarrow}$ is dramatic:
conductance is spin-unpolarized at negative bias and significantly
spin-polarized at positive bias. $TMR$ also changes significantly
around zero bias voltage, as shown in Fig.~\ref{fig3}-C and D.
Circles represent $TMR$ from Eq.~\ref{tmr}, and squares are
obtained as $TMR=\int G_{\uparrow\uparrow}(V)dV/\int
G_{\downarrow\uparrow}(V)dV-1$.

TMR values in our devices are less than 10\%. In state of the art
magnetic tunnelling junctions, $TMR$ exceeds 40\% at room
temperature and it is critically dependent upon the fabrication
process  and annealing of the tunnelling junctions.~\cite{parkin}
As discussed before, tunnelling in our samples is dominated by
weak spots caused by the surface roughness and oxygen vacancies.
The junctions are not ideal and thus the TMR is reduced.

In non-ideal magnetic tunnelling junctions, $TMR$ can be strongly
asymmetric and bias voltage
dependent.~\cite{zhang,tsymbal,tsymbal1,petta2} The asymmetry has
been explained by the two-step tunnelling via localized states. In
our control samples (without grains), $TMR$ is found to be
symmetric and weakly dependent on $V$.
This shows that the localized states responsible for asymmetric
$TMR$ in junctions with grains are the electronic states in the
grains.

We explain the asymmetry in $TMR$ by spin-relaxation in Al grains
and the asymmetry between the resistance between the grains and
the two reservoirs. Asymmetric resistances are easily introduced
by sample fabrication. For example, exposure of the bottom
aluminum oxide layer to air increases the oxide thickness by
hydration.

The average dwell time is $\tau_D\sim
\frac{R_D}{R_Q}\frac{h}{\delta}$, where $R_D$ is the average
resistance between the grains and the drain reservoir, and
$R_Q=26k\Omega$. When the bias voltage changes sign, the drain
reservoir changes, so $\tau_D$ also changes.

Asymmetric $TMR$ occurs when the spin-relaxation time $T_1$ is
smaller than the longer dwell time. For example, in sample 2,
figure~\ref{fig3}-B suggests that $T_1$ is much smaller than the
dwell time at negative bias, and $T_1$ is comparable to or longer
than the dwell time at positive bias. The voltage interval around
zero bias where the dwell times change is given by the Coulomb
Blockade thermal width $7k_BT/e$, in agreement with
Fig.~\ref{fig3}.

At large magnitude of bias voltage, $TMR$ has a complex dependence
on the magnitude of bias voltage. It is difficult to explain the
origin of this dependence, but we speculate that energy dependence
of the spin relaxation time, single electron charging effects, and
the
 distribution of energy level spacings in the ensemble of
 grains may play important
roles.

\begin{figure}
\includegraphics[width=0.47\textwidth]{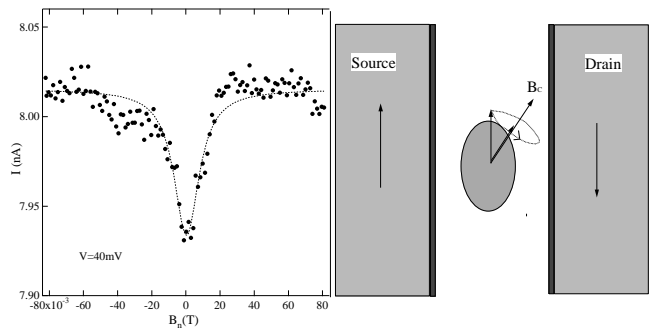}
\caption{A. Current versus perpendicular field in sample 3. B.
Sketch of the effects of the local field on TMR. The component of
the injected spin along the local field direction gives rise to a
finite TMR, even if $T_2^\star$ is much shorter than the dwell
time.~\label{fig4}}
\end{figure}

Next, we discuss the effects of spin precession in the applied
magnetic field (the Hanle effect). We measure current versus
magnetic field $B_n$ applied perpendicular to the film.
Fig.~\ref{fig4}-A displays the resulting peak in current versus
$B_n$, for sample 3 in the antiparallel configuration (in zero
applied parallel field). The dependence is reversible when $B_n$
is swept up and down, which shows that the curve does not arise
from the hysteresis loop in the leads. The peak amplitude is
$(I_{\uparrow\uparrow}-I_{\uparrow\downarrow})/2$.

The characteristic field $B_C$, defined as half-width of the peak,
is 8mT. We find that  $B_C$ is symmetric with bias voltage, which
shows that $B_C$ is independent of the dwell time. So, the
processes that contribute to the Hanle effect half-width are
different from the processes responsible for the $TMR$-asymmetry.
We have confirmed the Hanle effect in one more sample, however the
half-width was $2mT$.

The Hanle effect in a quantum dot has recently been calculated by
Braun et al.~\cite{braun} The calculation shows that perpendicular
field induces Larmour precession of the injected spin, which
reduces spin polarization of the current. Current versus $B_n$
exhibits a Lorentzian peak of amplitude
$(I_{\uparrow\uparrow}-I_{\uparrow\downarrow})/2$ (in agreement
with our data) centered at $B_n=0$. If a constant large parallel
magnetic field $B^\star$ is present, then the peak width becomes
$B^\star$ and the Hanle effect half-width is symmetric with bias
voltage.

Our observations (Fig.~\ref{fig4}-A) can be explained by these
theoretical results, if in zero applied field there exist a local
magnetic field $B_C$ acting on the grains. This local field
induces spin precession in zero applied magnetic field, and the
spin-coherence time is the Larmour period in the local field:
$T_2^\star\sim h/\mu_B B^\star \sim ns$.

The local field could be caused by the surface roughness, which
generates a finite dipole field originating from Co. Note that the
top aluminum-oxide/cobalt interface in Fig.~\ref{fig1}-B is quite
rough because of the underlying Aluminum grains. The local field
of 8mT is certainly possible because the internal field in Co is
2T. The hyperfine field from the nuclei can also create an
effective field of order mT that causes dephasing.~\cite{johnson1}
In our junctions, tunnelling is dominated by weak spots.
Consequently,  the local magnetic field will fluctuate among
samples, explaining the difference in $B_C$ between the samples.

$TMR$ survives dephasing because of the conservation of
spin-component along the local field direction. Even if the
electron dwell time is much longer than the dephasing time, $TMR$
will be finite. This is sketched in Fig.~\ref{fig4}-B. If the
magnetizations switch from parallel to antiparallel state at
finite $V$, then the injected spin component along the local field
direction switch from zero to finite value, giving rise to a
finite TMR. The perpendicular component of the injected spin is
averaged to zero, which reduces the TMR.

In conclusion, we study spin-polarized current through ensembles
of nm-scale Al grains.  The spin-coherence time is obtained from
the Hanle effect measurement: $T_2^\star \sim ns$. Fast dephasing
is attributed to spin-precession in the local magnetic field.
Tunnelling magnetoresistance is asymmetric with current direction,
which we attribute to the asymmetry in electron dwell times and
spin-relaxation. This work was performed in part at the
Georgia-Tech electron microscopy facility. This research is
supported by the David and Lucile Packard Foundation grant
2000-13874 and Nanoscience/Nanoengineering Research Program at
Georgia-Tech.

\bibliography{career1}

\end{document}